\documentclass[
]{ceurart}

\usepackage{listings}
\usepackage{float} 
\begin{document}

\conference{SIGIR eCom'25: Workshop on e-Commerce Search and Information Retrieval, July 2025, Padua, Italy}

\copyrightyear{2025}
\copyrightclause{Copyright for this paper by its authors. Use permitted under Creative Commons License Attribution 4.0 International (CC BY 4.0)}




\title{A Chain-of-Thought Approach to Semantic Query Categorization in e-Commerce Taxonomies}

\author[1]{Jetlir Duraj}[%
email=jduraj@ebay.com,
url=https://ceur-ws.org/Vol-4123/paper_24.pdf
]




\author[1]{Ishita Khan}[email=ishikhan@ebay.com]
\author[2]{Kilian Merkelbach}[%
email=kmerkelbach@ebay.com
]

\author[1]{Mehran Elyasi}[%
email=melyasi@ebay.com
]

\address[1]{eBay Inc, USA}

\address[2]{eBay Inc, Germany}



\begin{abstract}
Search in e-Commerce is powered at the core by a structured representation of the inventory, often formulated as a category taxonomy.
An important capability in e-Commerce with hierarchical taxonomies is to select a set of relevant leaf categories that are semantically aligned with a given user query.
In this scope, we address a fundamental problem of search query categorization in real-world e-Commerce taxonomies.
A correct categorization of a query not only provides a way to zoom into the correct inventory space, but opens the door to multiple intent understanding capabilities for a query.
A practical and accurate solution to this problem has many applications in e-commerce, including constraining retrieved items and improving the relevance of the search results.

For this task, we explore a novel Chain-of-Thought (CoT) paradigm that combines simple tree-search with LLM semantic scoring.
Assessing its classification performance on human-judged query-category pairs, relevance tests, and LLM-based reference methods, we find that the CoT approach performs better than a benchmark that uses embedding-based query category predictions.
We show how the CoT approach can detect problems within a hierarchical taxonomy.
Finally, we also propose LLM-based approaches for query-categorization of the same spirit, but which scale better at the range of millions of queries.   
\end{abstract}


\maketitle

\section{Introduction}\label{SEC:INTRO}


Mapping user queries to relevant categories is essential in e-Commerce search and navigation, since it enhances search relevance, user navigation, and inventory targeting. Traditionally, demand-based methods leveraging user behavioral data like click-through rates have been researched in industry and academic literature. However, these methods face issues like presentation bias and signal sparsity, particularly with long-tail queries and new inventory (e.g., \citet{clickthroughImplicit}, \citet{ecommerceSearchCollaborativeGraph}).

Recent semantic-based methods offer promising solutions by using linguistic and contextual understanding to infer relevance, addressing sparsity and bias while generalizing to new scenarios. These methods integrate query semantics with taxonomies for more precise mappings, but often lack task-specific adaptations and focus on static representations (e.g., \citet{neuralRankingWeak}, \citet{Gao2014ModelingIW}).

We propose a novel semantic projection system that complements demand-based methods. We adapt the chain-of-thought (CoT) reasoning paradigm for large language models (LLM) (see \citet{wei2022chain}) to our specific problem of classification in a hierarchical taxonomy. Given a query our system navigates from root to leaf categories of the taxonomy, integrating query semantics and taxonomy details to create precise, interpretable mappings. Our approach is orthogonal to demand-based methods and aims to enrich and de-bias them.

Similar in spirit to chain-of-thought (CoT) reasoning where an LLM solves a complex task by breaking it down into smaller steps and tasks, the system we build operates through a structured, multi-step process towards the solution. It iteratively predicts ranked categories at each taxonomy level, moving from root to leaf categories. Our model dynamically adjusts prediction thresholds based on the semantic information of the current category node, its children, and the query semantics.

A key feature of our approach is its ability to specify context for the query — such as user intent of buying, browsing, or accessory/complementary intents. Context specification enables more targeted and contextually appropriate category mapping. Our model provides confidence scores for each prediction, thus ranking categories with the aim of offering actionable insights and interpretability. Additionally, it can serve as a diagnostic tool for refining and improving taxonomies, addressing structural noise often misaligned with buyer signals.




\begin{figure*}[h]
  \centering
  \includegraphics[width=0.75\linewidth]{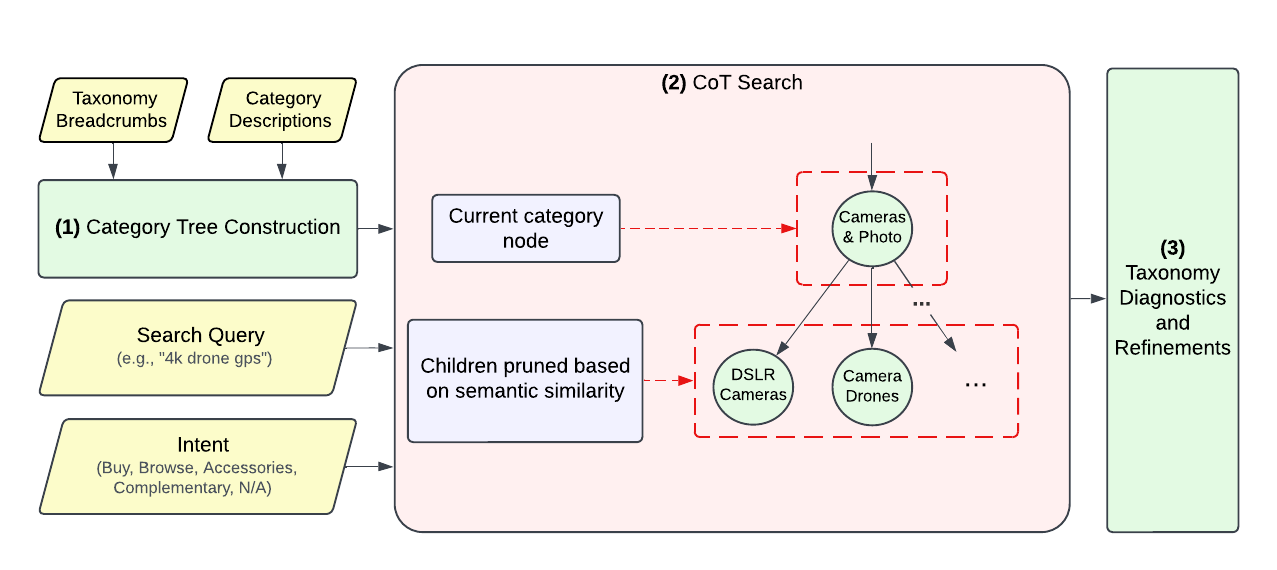}
  \caption{Overview of our system architecture. First, we build our category tree and enrich it with category descriptions. Then, we search within the tree for the node that best fits the search query. Finally, diagnostics and refinements of our taxonomy can be extracted.}
  \label{fig:architecture_overview}
\end{figure*}

\subsection{Related Work}

Recommendation systems and personalized search in e-Commerce heavily depend on understanding user interaction data semantically. Two main approaches exist in the literature: demand-based methods, using behavioral data, and semantic methods, enhancing recommendations through query and content understanding.

\paragraph{Demand-based approaches to category prediction} These approaches use implicit feedback, like click-through data, to personalize recommendations. \citet{deepListwiseRanking} introduced a model using local contexts for improved ranking, while \citet{clickthroughImplicit} highlighted click-through data's utility despite biases. \citet{ecommerceSearchCollaborativeGraph} used graph neural networks to better handle long-tail queries and cold-start products, thus tackling the persisting challenge of data sparsity.

\paragraph{Semantic-based methods and query understanding} The focus here is on understanding query intent and matching it with relevant content. \citet{deepRelevanceAdHoc} proposed a deep relevance matching model, and \citet{introToIR} introduced semantic embeddings for aligning queries with documents. These methods help mitigate bias and sparsity issues. \citet{neuralRankingWeak} demonstrated weak supervision's effectiveness in sparse datasets, while \citet{Gao2014ModelingIW} explored semantic generalization in taxonomies.

\paragraph{LLMs} LLMs are good at generalizing to unseen scenarios, which is a crucial capability for handling long-tail queries in e-Commerce. \citet{llmsFewShotLearners_NEURIPS2020_1457c0d6} showed LLMs' strengths in few-shot learning. Chain-of-thought reasoning (\citet{wei2023chainofthoughtpromptingelicitsreasoning, Kojima2022LargeLM}) enhances hierarchical reasoning and predictions.

\paragraph{Taxonomy integration} Taxonomies offer a structured basis for improving category projections and recommendations. \citet{deepStructuredModelsSemanticClickthrough} showed how to integrate semantic signals and taxonomies in e-Commerce search.

\section{Methodology}\label{sec:methodology}

\subsection{High level overview}


The task of identifying semantically relevant leaf categories is a multi-label classification problem given input data. The labels correspond to the leaf categories of a tree-structured taxonomy. This makes tree path-finding algorithms a natural choice to study. Additionally, assessing the strength of the semantic relationship between a query and leaf categories is crucial for applications. Therefore, we integrate straightforward tree search methods with LLM-scoring for semantic relevance. In LLM-scoring, the LLM is asked to provide a score measuring the strength of the semantic relation between a query and an e-commerce category and its description. LLM-scoring for semantic relevance closely approximates human judgment in our task, as demonstrated in Table \ref{tab:llm_classification_performance}.

\begin{table}[h]
    \caption{Classification performance of LLM scoring (human judgment as ground truth).}
    \label{tab:llm_classification_performance}
    \begin{tabular}{l|ccc}
        \toprule
      LLM  & F1 & Precision & Recall \\
        \midrule
        Mixtral-8x7B & 0.727 & 0.825 & 0.649 \\
        Llama3-70B   & \textbf{0.805} & 0.797 & \textbf{0.813} \\
        OpenAI-GPT-4oMini  & 0.743 & \textbf{0.862} & 0.654 \\
        \bottomrule
    \end{tabular}
\end{table}

This table presents classification metrics for the semantic relevance of 4,897 query-leaf category pairs, where human judges determine the ground truth. The LLMs we evaluated perform well in terms of F1, precision, and recall. In terms of inference speed, Mixtral-8x7B, a mixture-of-experts model that can be hosted locally, while ranked third in terms of F1, has significantly higher inference speed than the other two models considered.\footnote{The experiments took less than two hours to run within eBay's infrastructure. OpenAI-GPT-4oMini is closed source, while Llama3-70B is internally hosted.} We use Mixtral-8x7B for the results of our methodology in the paper.

Another reason for favoring LLM-scoring for semantic relevance over more generative LLM approaches is the issue of instruction-following. Such issues cannot be entirely eliminated due to the generative nature of current LLM architectures. When prompted to select directly from available children in a category node, LLMs sometimes modify category names rather than reproduce the exact names from inputs. This issue is less prevalent in closed-source LLMs like those from OpenAI or Gemini (4\% failure rate in our experiments), but is more pronounced in open-source models from Hugging Face, which can be hosted locally and support large-scale inference.

For these reasons, we focused on a scoring approach for the LLM component of our method: we ask the LLM for a confidence score for semantic relevance, ranging from 1 (lowest) to 10 (highest). The decision to continue searching at each category node is based on the semantic scores of its children, along with other contextual, or algorithm-runtime information.



\subsection{Breadth first search design (CoT BFS)}

Interpreting the categorization task as one of regression (score assignment) after classification, our approach first solves the classification problem in terms of relevant leaf categories fully, before addressing the regression problem. 
Specifically, for a given query, at level 1 of the taxonomy, after scoring the semantic relevance of all level-1 categories, we retain only the relatively most semantically relevant children, pruning the rest. We use two query-dependent thresholds for selection: a selection-threshold and a minimum-threshold, both ranging from 1 to 10. The relevance scores (1 to 10) of category children are mapped to the standard normal distribution. The selection-threshold divided by 10 is applied to the standardized scores to prune less relevant children. For example, with a selection-threshold of 9 (out of 10), children scoring below the mean plus 0.9 times the standard deviation of semantic scores are pruned. To deal with potential high skewness of the score distribution at the lower end, a child's original semantic score must exceed the minimum-threshold (the second threshold) to survive for further exploration.

Next, the algorithm examines each subtree starting from the level-1 children that survived the initial pruning. We repeat the pruning process for children in these subtrees to identify non-pruned second-level nodes using the same relative thresholding procedure described above. This iterative process continues until we reach nodes without children, which are leaf categories, and thus added to the set of candidate leaf categories.

Finally, because the search relies on relative rather than absolute semantic thresholding, we score the final set of leaf categories using only leaf category information (categorization path and descriptions). The surviving leaf categories with high semantic relevance, above the minimum-threshold, are the final predictions.

For our empirical application we choose the selection-threshold and minimum-threshold as follows: given the range of semantic scoring between 1 (lowest) and 10 (highest), we never consider thresholds below 6. Among thresholds 7, 8, 9, for both selection-threshold and minimum-threshold, we only look at pairs where selection-threshold is above the minimum-threshold. Among such pairs, we pick the one that does best in terms of F1-score against a human judgment dataset composed of about 1000 representative queries.\footnote{Note that the maximal score of 10, i.e. the span of the range 1..10 is also a hyperparameter. These hyperparameters need to be validated periodically over time, to account for distribution shifts of the queries, but also changes in the taxonomy. Space constraints preclude us from including a detailed analysis of the effect of the hyperparameters. Here we report qualitatively the following: lowering the selection-threshold and minimum-thresholds typically increases recall, but lowers precision. We also found that using the alternative range 1..5 instead of 1..10 lowered both precision and recall, while using 1..20 resulted in slightly higher recall, lower precision and lower F1.}

We refer to this method as the Chain-Of-Thought Breadth-First-Search (CoT BFS). 

\begin{figure}
    \centering
        \includegraphics[width=0.65\textwidth]{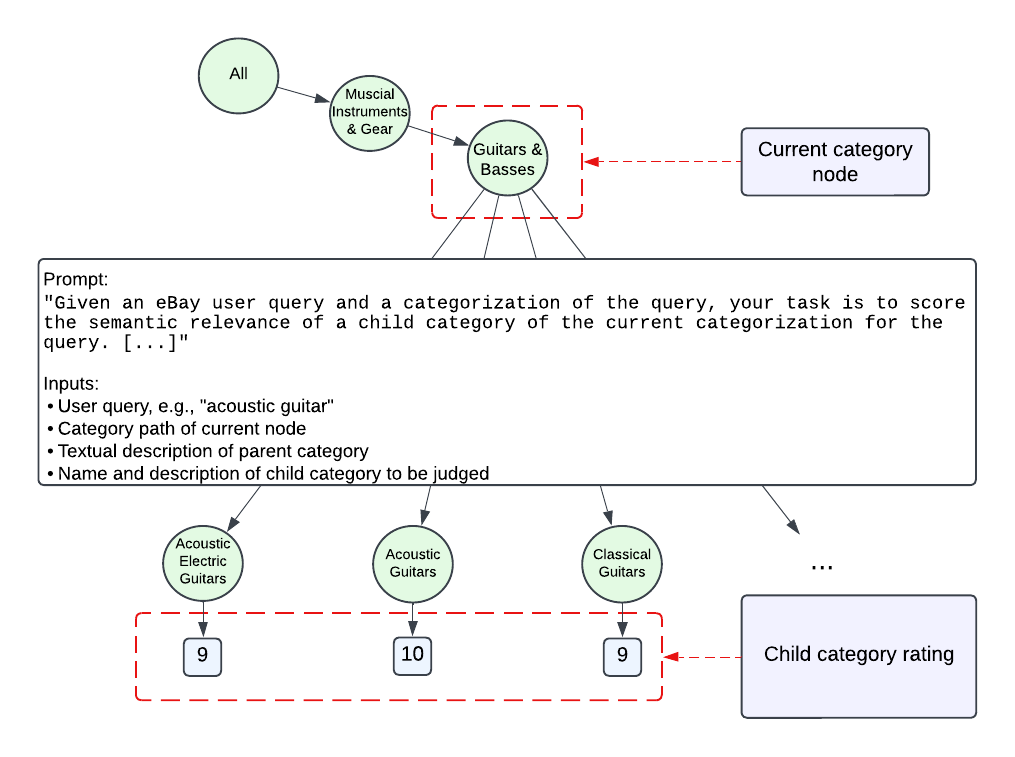}

    \caption{CoT BFS output for the query \emph{acoustic guitar}. Mixtral-8x7B was used as the LLM. The prompt contains information about the path from the root of the tree to the current node, the immediate parent node, and the child node to be rated.}
    \label{fig:bfs_tree}
\end{figure} 

Figure \ref{fig:bfs_tree} illustrates the CoT BFS categorization result for the query \emph{acoustic guitar}, with selection-threshold of 9, and minimum-threshold of 8. In the first step, CoT BFS narrows down to a single level-1 category: \emph{AllCats \textgreater{} Musical Instruments \& Gear}, which gets an intermediate semantic score of 10. Out of 35 level-1 categories all other level 1 categories have low score, with the mode score of 1, and maximal other score of 4 (category \emph{AllCats \textgreater{} Music}). The surviving level-1 category \emph{AllCats \textgreater{} Musical Instruments \& Gear} has 16 children. The next classification step prunes out 15 out of these 16 children, because all of them have semantic score lower than 8. The child \emph{Guitars \& Basses} has semantic score of 10 and survives. Further, the node \emph{AllCats \textgreater{} Musical Instruments \& Gear \textgreater{} Guitars \& Basses} has 13 children. Scoring these children in the next step  prunes out all but three children. The surviving children are the nodes:
\emph{Classical Guitars} (with a final score of 9), \emph{Acoustic Electric Guitars} (final score of 9), and \emph{Acoustic Guitars} (final score of 10). At this point, the search stops, as the reached nodes are already leaf-categories of the category tree.

\subsection{Scalable approaches for CoT and LLM scoring}\label{subsec:scalable}


For a given query, the total number of LLM calls in CoT BFS is in the same order with the number of category nodes visited in the taxonomy.
Experiments on large datasets of queries show CoT BFS can visit between 1.7\% to 24.8\% of the total number of category nodes of eBay's taxonomy.
This shows the efficiency of our method, given the very high number of categories in eBay's taxonomy.
Nonetheless, to scale this method to millions of queries and low latency, modifications are needed. We propose two approaches, the second more scalable than the first.


\subsubsection{CoT-k-NN hybrid BFS}

k-NN retrieval based on embeddings of category names or descriptions can be used as a filter at each step of the tree search process.
Instead of exhaustively rating each child node, only a subset surviving the embedding distance filter (between the user query and a textual representation of the category) is scored by the LLM.
This reduces the number of LLM calls at each node of the taxonomy, and constrains the search to only the most promising directions.


\subsubsection{k-NN-search + LLM scoring on leaf categories}\label{subsubsec:hybridmethod} One replaces tree-search with a k-NN-search on leaf category embeddings as a pre-filter, followed by LLM scoring of the candidates identified through k-NN. Running k-NN with many neighbors at the beginning of the procedure, e.g. 20 neighbors, enhances recall. We use a variant of this method in section \ref{subsec:eval} to construct a synthetic ground truth for evaluating CoT BFS. 

\section{Experimentation}
\subsection{Baseline model: k-nearest neighbors categorization}

Our benchmark for evaluation of the CoT BFS is k-NN search for leaf categories with $k=10$ using (not fine-tuned) embeddings from sentence-BERT. Cosine-similarity is used as a metric for the k-NN.\footnote{We have access to language models trained on eBay-specific data, that typically perform better in eBay-related tasks than general-purpose language models, but we do not present results from the use of eBay-specific language models. This is because our focus is on understanding how the CoT BFS approach performs with general-purpose, non-fine-tuned LLMs. Furthermore, using language models that are publicly accessible helps with the reproducibility of the results.} We provide detailed categorization performance, comparing our method's F1, precision and recall classification metrics in the micro, macro and sample aggregations. Micro aggregation considers performance across all queries and leaf categories. Sample aggregation considers performance per-query and then aggregates. Macro aggregation considers performance per leaf category and then aggregates.

\subsection{Evaluation against baseline}\label{subsec:eval}


\subsubsection{Human Judgment}  

Human judgment offers both qualitative and quantitative evaluations by utilizing human intuition and expertise. Evaluators review predicted categories for semantic relevance, though this process is subjective and costly for large datasets. Despite its costs, human judgment captures nuances often missed by other methods, hence it is indispensable. Our human judgment dataset includes 1018 queries and 4897 query-category pairs, judged on semantic relevance (Yes/No decision). We note that the leaf categories for judgment were chosen based on user behavior signals, leading to presentation bias influenced by eBay's current models in production. The annotators are three eBay-funded independent domain experts for eBay's taxonomy. 

Table \ref{tab:metrics_hj_relative} shows the relative performance of the CoT BFS to the benchmark, assuming that the ground truth is given by the human judgment. CoT BFS outperforms the baseline model, especially in relation to the F1 score and precision. 

\begin{center}
\begin{table}[h]
\caption{Classification Results on Human judgment data. CoT BFS Select-threshold 9, min-threshold 8. Relative improvement to benchmark.}
\label{tab:metrics_hj_relative}
\begin{tabular}{l|ccc}
\toprule
Metric & F1 & Precision & Recall \\
\midrule
Micro Aggregation & +89.8\% & +86.1\% & +3.4\% \\
\midrule
Sample aggregation & +109.7\% & +190.6\% & -13.5\%\\
\midrule
Macro aggregation & +44.7\% & +60\% & +31.4\%\\
\bottomrule
\end{tabular}
\end{table}

\end{center}

\subsubsection{AI Pseudo-Reference Method}  

We use an AI pseudo-reference method to create a dataset that approximates ground truth without the presentation bias found in demand signal datasets. Starting with 3000 user queries, a high-quality LLM emulates human judgment on semantic relevance for query-category pairs. To avoid losing potentially relevant leaf categories for LLM-scoring, we use a k-NN embedding-based search with a large number of neighbors. Namely, we pick out the 100 most relevant categories for each query. From these, we exclude those with cosine similarity below 0.01. 

Afterwards, each pair is scored from 1 to 10 using a superior LLM (OpenAI-GPT-4o-Mini) compared to locally hosted Mixtral-8x7B we use for CoT BFS. This hybrid method with large $k$, see also subsection \ref{subsubsec:hybridmethod}, delivers a proxy for ground truth. Table \ref{tab:metrics_random_3k_relative} depicts the results. CoT BFS again outperforms the baseline, especially in terms of F1 and precision. 
\begin{table}[h]
\centering
\caption{Classification Results on LLM-judged Data (OpenAI-GPT-4oMini). CoT BFS Select-threshold 9, min-threshold 8. Relative improvement to benchmark.}
\label{tab:metrics_random_3k_relative}
\begin{tabular}{l|ccc}
\toprule
Metric & F1 & Precision & Recall \\
\midrule
Micro Aggregation & +96.7\% & +137.5\% & -5.4\% \\
\midrule
Sample aggregation & +132.1\% & +89.8\% & +3.7\%\\
\midrule
Macro aggregation & +112.1\% & +47.8\% & +111.4\%\\
\bottomrule
\end{tabular}
\end{table}

\subsubsection{Retrieval Test}  

The retrieval test evaluates predicted leaf categories by comparing recall and relevance  between our model and the baseline at the level of retrieved items. Items from the inventory are retrieved based on leaf categories, with estimated recall size showing the proportion of relevant items found. Relevance is measured using an eBay-internal PEGFB model that has been trained on human judgment data, and which classifies results into five graded relevance levels: Perfect, Excellent, Good, Fair, and Bad.  The retrieval test evaluation highlights the model's practical utility in improving user satisfaction and search efficiency. Our model significantly outperforms the k-NN benchmark in both recall and relevance, with Mann-Whitney U test results showing highly significant differences in favor of CoT BFS.
\vspace{-2mm}
\begin{table}[h]
\caption{Classification Results for 3000 queries. CoT BFS Select-threshold 9, min-threshold 8. Relative improvement to benchmark.}
\label{tab:metrics_retrieval_test}
\begin{tabular}{l|cc}
\toprule
 & Estimated Recall Size & Relevance Score \\
\midrule
Mean & +72\% & +34\% \\
\midrule
Mann-Whitney U stat & 5401539.0 & 5363320.0\\
\midrule
Mann-Whitney U p-value & 8.95e-88 & 1.30e-82\\
\bottomrule
\end{tabular}
\end{table}

\vspace{-2mm}
\section{Applications}
\subsection{Context learning}

 By learning from extensive real-world datasets, LLMs can identify patterns that reveal user intent and preferences, enabling personalized search leading to higher semantic relevance. Contextual learning can refine model outputs based on a given context of the query, ensuring that prediction results are relevant. This capability is important for platforms like eBay, where discerning buyer intent enhances the search experience. We consider two applications of context learning, one on user intent and one on brand origin. 

\begin{figure}[h]
    \centering
        \includegraphics[width=0.6\textwidth]{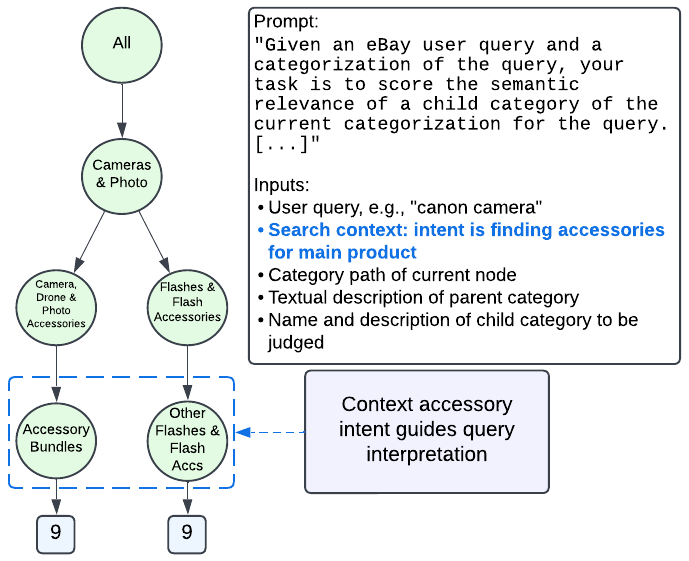}
    \caption{CoT BFS output for the query \emph{canon camera} with accessory intent. The query intent being supplied as context guides the category search into categories that correspond to accessories for the entity in the query.}
    \label{fig:fig_intent_tree}
\end{figure}

More specifically, eBay's taxonomy includes accessory-related categories across various merchandise segments like electronics, automotive, and fashion. The CoT BFS approach can easily incorporate buyer intent by modifying LLM prompts to include intents such as \emph{buying}, \emph{seeking accessories}, \emph{looking for complementary items}. Figure~\ref{fig:fig_intent_tree} illustrates for the query \emph{canon camera}.

Without intent, the top category identified for the query is \emph{AllCats \textgreater{} Cameras \& Photo \textgreater{} Digital Cameras} with a score of 10. By injecting \emph{accessory} intent as a search context into the prompt, the top categories identified are \emph{AllCats \textgreater{} Cameras \& Photo \textgreater{} 
 Camera, Drone \& Photo Accessories \textgreater{} Accessory Bundles} and \emph{AllCats \textgreater{} Cameras \& Photo \textgreater{} Flashes \& Flash Accessories \textgreater{} Other Flashes \& Flash Accs}, with a score of 9 each. More generally, table \ref{tab:accessories} shows how the average semantic scores for Accessory vs. No-Accessory predicted categories change for 15 selected queries, when specifying \emph{accessory} intent.

\begin{table}[h]
\caption{Context learning: Accessory Intent Improves Prediction for Accessory Categories.}
\label{tab:accessories}
\begin{tabular}{l|ccc}
\toprule
Acc. Category ? (Y/N) & Avg. Score for No Intent & Avg. Score for Acc. Intent \\
\midrule
No & \textbf{9.143} & 1.000 \\
Yes & 2.643 & \textbf{7.214} \\
\bottomrule
\end{tabular}
\end{table}

Including buyer intent in CoT BFS leads to better targeting of relevant categories.

Similarly, we consider the effect of injecting context regarding brand origin. To illustrate, for the query \emph{sports car}, we consider the two distinct contexts of \emph{brand origin is from Germany}, and \emph{brand origin is from Italy}. For this query, the CoT BFS predicts exclusively sports car-related leaf categories from the german brands Audi, BMW, Mercedes-Benz, Porsche in the first case, and exclusively leaf categories from the italian brands Alfa Romeo, De Tomaso, Ferrari, Fiat, Maserati, Lamborghini in the second. 

\subsection{Detecting issues and improving the taxonomy}


By analyzing query patterns and identifying gaps in category representation, CoT BFS can help provide actionable insights for improving e-Commerce taxonomy structures.

In this regard, we conducted an experiment using a representative sample of 25,000 queries processed through the CoT BFS approach with high thresholds: selection-threshold of 10, minimum-threshold of 9. There were 3,110 queries with empty model predictions, indicating that the current eBay taxonomy lacks category nodes at the first few levels that are strongly semantically related to these queries. We uncovered certain patterns by clustering these queries using k-NN search on embeddings. For instance, two clusters of these "failing" queries correspond to the e-Commerce categories \emph{Designer Sunglasses} and \emph{Optical Instruments and Accessories}. The closest leaf categories in the current eBay taxonomy for the first cluster identified (\emph{Designer Sunglasses}) are \emph{AllCats \textgreater{} Clothing, Shoes \& Accessories \textgreater{} Women \textgreater{} Women's Accessories \textgreater{} Sunglasses \& Sunglasses Accessories \textgreater{} Sunglasses} and \emph{AllCats \textgreater{} Clothing, Shoes \& Accessories \textgreater{} Men \textgreater{} Men's Accessories \textgreater{} Sunglasses \& Sunglasses Accessories \textgreater{} Sunglasses}, both in a depth of 5 in the taxonomy. A similar issue is observed for the second cluster. These types of insights, when drawn from large sets of user queries, can help product management teams in their taxonomy enhancements work. E.g., introducing a level-2 category titled \emph{Optical Products and Eyewear} with subcategories such as \emph{Designer Sunglasses} and \emph{Optical Instruments and Accessories} might be beneficial for the search experience.

Ultimately, maintaining an e-Commerce taxonomy that provides high value to users involves complex business and product management decisions. Our methodology offers tools to explore taxonomy issues with the goal of enhancing decision making in these complex business decisions.

\section{Conclusion and future work}

In this study, we introduce a novel methodology for query categorization within hierarchical taxonomies. It combines the world knowledge of LLMs and simple tree search algorithms to achieve high-quality categorization and provide deep insights into the taxonomy.

AB-tests are planned for the scalable methods presented in section \ref{subsec:scalable}. These involve direct tests where the model predictions are cached for use in production, but also indirect tests where the AB test is on lower-latency categorization models trained with data that have been LLM-labeled via CoT BFS.  

Further, we developed a CoT algorithm version that uses absolute thresholding at each taxonomy node, rather than the relative thresholding discussed in the paper, here left out due to space constraints. This method, called Chain-of-Thought Depth-first-search (CoT DFS), searches for leaf categories in a depth-first manner and halts a path when encountering an intermediate node with low absolute semantic relevance, as opposed to relative described in this paper. Because of its more stringent requirements, the CoT DFS approach leads to more queries with empty predictions. CoT DFS can leverage user query activity and LLM semantic-knowledge more effectively than CoT BFS for the purpose of taxonomy diagnostics.

\section*{Declaration on Generative AI}
The author(s) have not employed any Generative AI tools.

\bibliography{references}

@misc{wei2023chainofthoughtpromptingelicitsreasoning,
      title="Chain-of-Thought Prompting Elicits Reasoning in Large Language Models", 
      author="Jason Wei and Xuezhi Wang and Dale Schuurmans and Maarten Bosma and Brian Ichter and Fei Xia and Ed Chi and Quoc Le and Denny Zhou",
      year="2023",
      eprint="2201.11903",
      archivePrefix="arXiv",
      primaryClass="cs.CL",
      url="https://arxiv.org/abs/2201.11903", 
}

@article{Kojima2022LargeLM,
  title="Large Language Models are Zero-Shot Reasoners",
  author="Takeshi Kojima and Shixiang Shane Gu and Machel Reid and Yutaka Matsuo and Yusuke Iwasawa",
  journal="ArXiv",
  year="2022",
  volume="abs/2205.11916",
  url="https://api.semanticscholar.org/CorpusID:249017743"
}

@inproceedings{deepStructuredModelsSemanticClickthrough,
author = "Huang, Po-Sen and He, Xiaodong and Gao, Jianfeng and Deng, Li and Acero, Alex and Heck, Larry",
title = "Learning deep structured semantic models for web search using clickthrough data",
year = "2013",
isbn = "9781450322638",
publisher = "Association for Computing Machinery",
address = "New York, NY, USA",
url = "https://doi.org/10.1145/2505515.2505665",
doi = "10.1145/2505515.2505665",
booktitle = "Proceedings of the 22nd ACM International Conference on Information \& Knowledge Management",
pages = "2333–2338",
numpages = "6",
location = "San Francisco, California, USA",
series = "CIKM '13"
}

@inproceedings{Gao2014ModelingIW,
  title="Modeling Interestingness with Deep Neural Networks",
  author="Jianfeng Gao and Patrick Pantel and Michael Gamon and Xiaodong He and Li Deng",
  booktitle="Conference on Empirical Methods in Natural Language Processing",
  year="2014",
  url="https://api.semanticscholar.org/CorpusID:2141094"
}

@inproceedings{llmsFewShotLearners_NEURIPS2020_1457c0d6,
 author = "Brown, Tom and Mann, Benjamin and Ryder, Nick and Subbiah, Melanie and Kaplan, Jared D and Dhariwal, Prafulla and Neelakantan, Arvind and Shyam, Pranav and Sastry, Girish and Askell, Amanda and Agarwal, Sandhini and Herbert-Voss, Ariel and Krueger, Gretchen and Henighan, Tom and Child, Rewon and Ramesh, Aditya and Ziegler, Daniel and Wu, Jeffrey and Winter, Clemens and Hesse, Chris and Chen, Mark and Sigler, Eric and Litwin, Mateusz and Gray, Scott and Chess, Benjamin and Clark, Jack and Berner, Christopher and McCandlish, Sam and Radford, Alec and Sutskever, Ilya and Amodei, Dario",
 booktitle = "Advances in Neural Information Processing Systems",
 editor = "H. Larochelle and M. Ranzato and R. Hadsell and M.F. Balcan and H. Lin",
 pages = "1877--1901",
 publisher = "Curran Associates, Inc.",
 title = "Language Models are Few-Shot Learners",
 url = "https://proceedings.neurips.cc/paper_files/paper/2020/file/1457c0d6bfcb4967418bfb8ac142f64a-Paper.pdf",
 volume = "33",
 year = "2020"
}

@inproceedings{neuralRankingWeak,
author = "Dehghani, Mostafa and Zamani, Hamed and Severyn, Aliaksei and Kamps, Jaap and Croft, W. Bruce",
title = "Neural Ranking Models with Weak Supervision",
year = "2017",
isbn = "9781450350228",
publisher = "Association for Computing Machinery",
address = "New York, NY, USA",
url = "https://doi.org/10.1145/3077136.3080832",
doi = "10.1145/3077136.3080832",
booktitle = "Proceedings of the 40th International ACM SIGIR Conference on Research and Development in Information Retrieval",
pages = "65–74",
numpages = "10",
location = "Shinjuku, Tokyo, Japan",
series = "SIGIR '17"
}

@article{introToIR,
url = "http://dx.doi.org/10.1561/1500000061",
year = "2018",
volume = "13",
journal = "Foundations and Trends® in Information Retrieval",
title = "An Introduction to Neural Information Retrieval",
doi = "10.1561/1500000061",
issn = "1554-0669",
number = "1",
pages = "1-126",
author = "Bhaskar Mitra and Nick Craswell"
}

@inproceedings{deepRelevanceAdHoc,
author = "Guo, Jiafeng and Fan, Yixing and Ai, Qingyao and Croft, W. Bruce",
title = "A Deep Relevance Matching Model for Ad-hoc Retrieval",
year = "2016",
isbn = "9781450340731",
publisher = "Association for Computing Machinery",
address = "New York, NY, USA",
url = "https://doi.org/10.1145/2983323.2983769",
doi = "10.1145/2983323.2983769",
booktitle = "Proceedings of the 25th ACM International on Conference on Information and Knowledge Management",
pages = "55–64",
numpages = "10",
location = "Indianapolis, Indiana, USA",
series = "CIKM '16"
}

@inproceedings{ecommerceSearchCollaborativeGraph,
author = "Xv, Guipeng and Lin, Chen and Guan, Wanxian and Gou, Jinping and Li, Xubin and Deng, Hongbo and Xu, Jian and Zheng, Bo",
title = "E-commerce Search via Content Collaborative Graph Neural Network",
year = "2023",
isbn = "9798400701030",
publisher = "Association for Computing Machinery",
address = "New York, NY, USA",
url = "https://doi.org/10.1145/3580305.3599320",
doi = "10.1145/3580305.3599320",
booktitle = "Proceedings of the 29th ACM SIGKDD Conference on Knowledge Discovery and Data Mining",
pages = "2885–2897",
numpages = "13",
location = "Long Beach, CA, USA",
series = "KDD '23"
}

@inproceedings{clickthroughImplicit,
author = "Joachims, Thorsten and Granka, Laura and Pan, Bing and Hembrooke, Helene and Gay, Geri",
title = "Accurately interpreting clickthrough data as implicit feedback",
year = "2005",
isbn = "1595930345",
publisher = "Association for Computing Machinery",
address = "New York, NY, USA",
url = "https://doi.org/10.1145/1076034.1076063",
doi = "10.1145/1076034.1076063",
booktitle = "Proceedings of the 28th Annual International ACM SIGIR Conference on Research and Development in Information Retrieval",
pages = "154–161",
numpages = "8",
location = "Salvador, Brazil",
series = "SIGIR '05"
}

@inproceedings{deepListwiseRanking,
author = "Ai, Qingyao and Bi, Keping and Guo, Jiafeng and Croft, W. Bruce",
title = "Learning a Deep Listwise Context Model for Ranking Refinement",
year = "2018",
isbn = "9781450356572",
publisher = "Association for Computing Machinery",
address = "New York, NY, USA",
url = "https://doi.org/10.1145/3209978.3209985",
doi = "10.1145/3209978.3209985",
booktitle = "The 41st International ACM SIGIR Conference on Research \& Development in Information Retrieval",
pages = "135–144",
numpages = "10",
location = "Ann Arbor, MI, USA",
series = "SIGIR '18"
}

@inproceedings{wei2022chain,
  author    = {Jason Wei and Xuezhi Wang and Dale Schuurmans and Maarten Bosma and Brian Ichter and Fei Xia and Ed H. Chi and Quoc V. Le and Denny Zhou},
  title     = {Chain-of-thought prompting elicits reasoning in large language models},
  booktitle = {Advances in Neural Information Processing Systems},
  volume    = {35},
  pages     = {24824--24837},
  year      = {2022},
}

\end{document}